\documentclass[aps,pre,preprint, showpacs, superscriptaddress]{revtex4}
\usepackage[dvips]{graphicx}
\usepackage{latexsym}
\usepackage{natbib,amsfonts}
\usepackage{epsfig}

\begin{document}
\title{Discrete surface growth process as a synchronization mechanism for 
  scale free complex networks}
\author{A. L. Pastore y Piontti}
\affiliation{Departamento de F\'{\i}sica, Facultad de Ciencias Exactas y
  Naturales, Universidad Nacional de Mar del Plata, Funes 3350, 7600 Mar del
  Plata, Argentina}
\author{P. A. Macri}
\affiliation{Departamento de F\'{\i}sica, Facultad de Ciencias Exactas y
  Naturales, Universidad Nacional de Mar del Plata, Funes 3350, 7600 Mar del
  Plata, Argentina}
\author{L. A. Braunstein}
\affiliation{Departamento de F\'{\i}sica, Facultad de Ciencias Exactas y
  Naturales, Universidad Nacional de Mar del Plata, Funes 3350, 7600 Mar del
  Plata, Argentina}
\affiliation{Center for polymer studies, Boston University, Boston, MA 02215, USA}

\begin{abstract}
We consider the discrete surface growth process with relaxation to the
minimum [F. Family, J. Phys. A {\bf 19} L441, (1986).] as a possible
synchronization mechanism on scale-free networks, characterized by a degree
distribution $P(k) \sim k^{-\lambda}$, where $k$ is the degree of a node and
$\lambda$ his broadness, and compare it with the usually applied
Edward-Wilkinson process(EW) [S. F. Edwards and D. R. Wilkinson,
Proc. R. Soc. London Ser. A {\bf 381},17 (1982) ].  In spite of both
processes belong to the same universality class for Euclidean lattices, in
this work we demonstrate that for scale-free networks with exponents
$\lambda<3$ the scaling behavior of the roughness in the saturation cannot be
explained by the EW process.  Moreover, we show that for these ubiquitous
cases the Edward-Wilkinson process enhances spontaneously the synchronization
when the system size is increased. This non-physical result is mainly due to
finite size effects due to the underlying network. Contrarily, the discrete
surface growth process do not present this flaw and is applicable for every
$\lambda$.
\end{abstract}

\pacs{89.75.Hc, 81.15.Aa, 68.35.Ct, 05.10.Gg}

\maketitle

The study of the dynamics on complex networks became a subject of
great interest in the last few years since it was realized that they
are useful tools to understand biological, social and communications
systems \cite{rmpBarabasi,DMO2}. Networks are constituted by nodes
associated to individuals, organizations or computers and by links
representing their interactions. The classical model for random
networks is the Erd\H{o}s-R\'enyi (ER) model
\cite{ER59,ER60,Bollobas} characterized by a Poisson degree
distribution $P(k) = \exp \left[-\langle k \rangle\right] \langle k
\rangle ^k/k!  $ where $k$ is the degree or number of links that a
node has and $\langle k \rangle$ is the average degree. However, it
was found \cite{rmpBarabasi} that many real networks are
characterized by a scale-free (SF) degree distribution given by
\begin{equation}\label{Eq.Pk}
P(k)= (\lambda-1)\; k_{min}^{\lambda-1} \frac{1}{1-(k_{min}/k_{max})^{\lambda-1}} k ^{- \lambda}\;,
\end{equation}
for $ k_{min} < k < k_{max}$ where $k_{min}$ is the smaller degree that a
node can have, $k_{max}$ is a cutoff that diverge when the system size
$N \to \infty$ and $\lambda$ represent the broadness of the
distribution. Most of observed networks such as Internet, the World Wide Web
and metabolic networks have $2 < \lambda <3$ \cite{rmpBarabasi,DMO2}.

It was shown that the topology of the network is very relevant to
determine their statics and dynamics properties, such as robustness
and percolation thresholds ~\cite{Cohen_robust, Callaway}, the
average shortest path length ~\cite{Cohen_ultrasmall} and
transport~\cite{Lopez_transport}. An important quantity
characterizing networks is its diameter (maximal hopping) $d$. In a
network of a total of $N$ nodes, $d$ scales as $\ln N$
\cite{Bollobas}, which leads to the concept of ``small worlds'' or
``six degrees of separation'' \cite{Watts} . For scale-free (SF)
networks with $\lambda <3$  \cite{Albert02} $d$ scales as $\ln \ln
N$, which leads to the concept of ultra small worlds
\cite{Cohen,DMO2}.

Very recently, the research focus is changing from the study of the
network topology to the study of dynamical processes on the
underlying network. Of particular interest are the studies on the
dynamics and fluctuations of task completion landscapes of queuing
networks. If for each node on the network there is a scalar $h$
which specifies the time it takes to finish a job or the amount of
work that has been assigned to it, the fluctuations on $h$ indicates
how synchronized or balanced is the system. Jobs synchronization and
load balance are required in many applications such as packet routing
on the Internet \cite{Valverde} or in parallel computing
\cite{Rabani98,novotny}.

These synchronization processes are usually mapped into a
non-equilibrium surface growth via an Edwards-Wilkinson (EW)
equation \cite{EW} on complex networks
\cite{Korniss06,Kozma04,Kozma05}. The EW equation for the evolution
of the growing interface in complex networks is given by

\begin{equation}\label{Eq.ew}
  \frac{\partial h_i}{\partial t}= \nu \sum_{j=1}^N A_{i j}  (h_j -h_i ) + \eta_i\;,
\end{equation}

where $h_i \equiv h_i(t)$ is the height of the interface of node
$i$, $A_{i j}$ is the element $i j$ of the adjacency matrix that
take the value $1$ if $i$ and $j$ are connected and zero otherwise,
$N$ is the system size, $\nu$ is a coefficient that represent the
``surface tension'' and $ \eta_i \equiv \eta_i(t)$ is a random
Gaussian uncorrelated noise with
  $\{ \eta_i\} =0$;
  $\{ \eta_i \eta_j\} = 2D \delta_{i j} \delta(t -t ^{'})$,
where $D$ is the diffusion coefficient and $ \{ . \}$ represent
averages over configurations. The interface is characterized by its
roughness $W(t)$ at time $t$, 
\begin{equation}\label{Eq.w} W(t) =
\left\{{\frac{1}{N} \sum_{i=1}^N (h_i - \langle h
\rangle)^2}\right\}^{1/2}\;,
\end{equation}
that represents the fluctuations of the height of the interface
around his mean value $ \langle h \rangle$.

There are several technical advantages of using the continuous EW equation to
model queue synchronization or load balance processes
\cite{Korniss06,Kozma04,Kozma05} mainly because it is a linear continuous equation. However, some
real implementations of this processes are intrinsically discrete. For this
reason, in this work we use the discrete growth model of surface relaxation
to the minimum (SRM), which is very well known on Euclidean lattices
\cite{family,barabasi}, on SF networks. It is also well known the fact that on
Euclidean lattices this discrete model belongs to the same universality class
of the EW equation. This might be one of the motivations of using EW to
model these discrete process. In the SRM model at each time step a node $i$
is chosen with probability $1/N$. If we denote by $v_i$ the nodes nearest
neighbors of $i$, then

\begin{eqnarray}\label{Eq.rules}
\textrm{if} \left \{\begin{array}{ll}
  h_i \leq h_j\; \forall j \in v_i  & \textrm{$\Rightarrow h_i= h_i+1$, else,} \nonumber\\
 \textrm{node $j$ has the minimum height $\in v_i$} &   \textrm{ $\Rightarrow h_j = h_j+1$}. \nonumber\\
\end{array}\right.\nonumber\\
\end{eqnarray}

This rules mimics a process where the higher loaded node distributes
the excess of load to one of his neighbors which is less charged. 
To generate SF graphs of size $N$, we employ the Molloy-Reed algorithm (MR)
\cite{Molloy}: initially the degree of each node is chosen according
to a SF distribution, where each node is given a number of open
links or "stubs" according to its degree. Then, stubs from all nodes
of the network are interconnected randomly to each other with the
two constraints that there are no multiple edges between two nodes
and that there are no looped edges with identical ends.

 We use for the simulation $k_{min}=2$ because
when $k_{min} > 1$ there is a high probability that the network is
fully connected \cite{Cohen} which is required in this work to
analyze the interface.

At $t=0$ we initialize all the values of $h_i$ with random numbers taken from an
uniform distribution in $[0,1]$.  At each time step we select a node with
probability $1/N$ and use the rules given by Eq.~(\ref{Eq.rules}), then the
time is increased by $1/N$. We compute $W(t)$ for SF networks with $\lambda
>2$ and different values of $N$.

\begin{figure}
\begin{center}
\includegraphics[width=8cm,height=10cm,angle=-90]{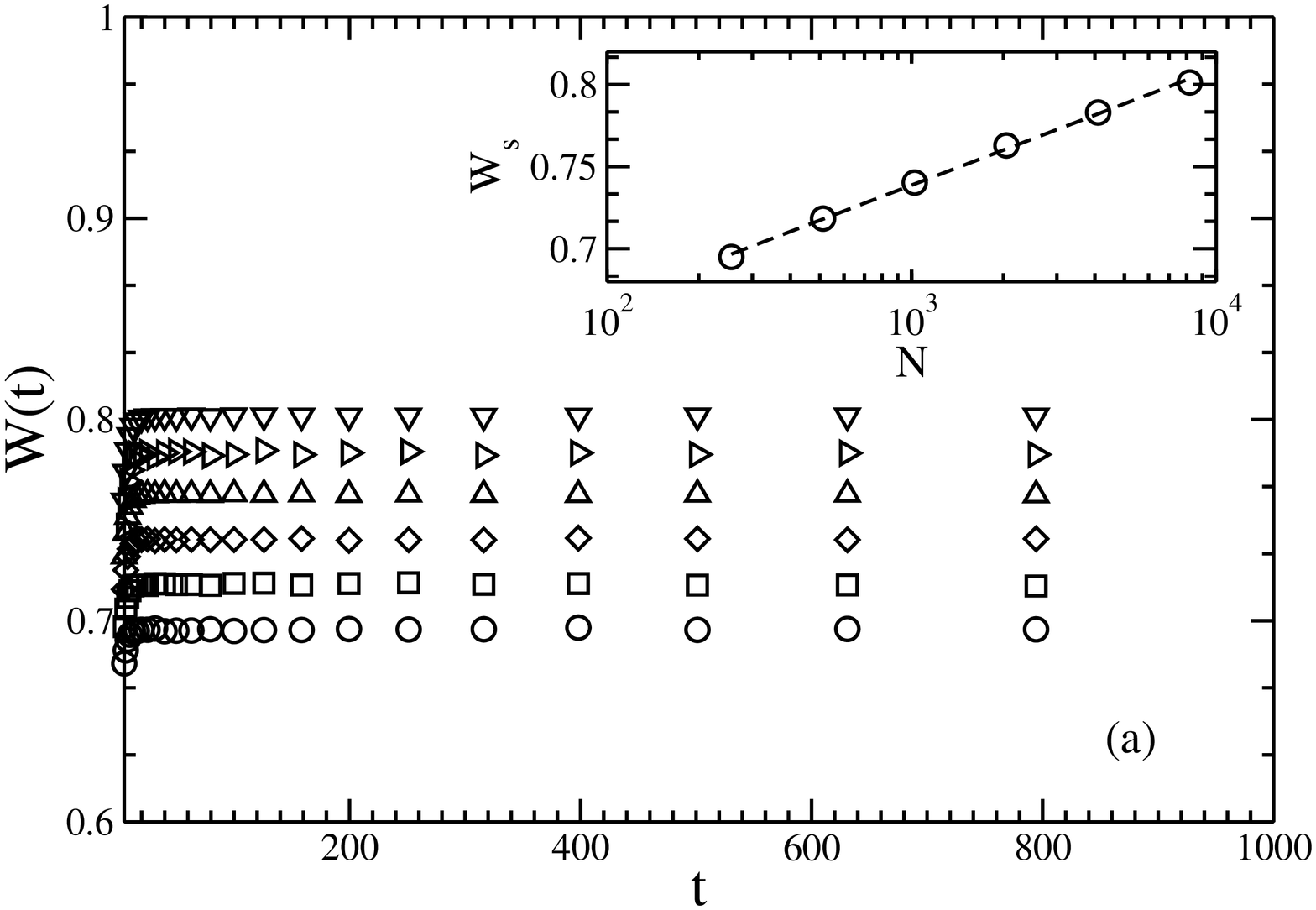}\\

\vspace{1cm}
\includegraphics[width=9cm,height=6cm,angle=0]{fig1b.eps}
\caption{Plots of $W(t)$ for SRM and different system size $N$,
$N=256$ ($\bigcirc$), $N=512$ ($\Box$), $N=1024$ ($\diamond$), $N=2048$
($\bigtriangleup$), $N=4096$ ($\triangleleft$) and $N=8192$
($\bigtriangledown$) for : (a) $\lambda=2.5$ we can see that $W_s$ increase with
the system size $N$. In the inset figure we show 
$W_s$ as function of $N$ in linear log scale ($\bigcirc$). The dashed lines
represent the  logarithmic fitting supporting that $W_s\sim \ln N$. 
(b) $\lambda=3.5$ we can see that $W(t)$ depend weakly on $N$. In the inset
figure we show   in symbols $W_s^2$ as function of $N$.
The dashed lines represent the fitting of $W_s^2$ with Eq.(\ref{Eq.wsn}) ($A\approx 10 $ and $B\approx 0.25 $). In all the inset of data's figures we do not display the errors bars because they are of the size of the symbols.\label{simulacion}}
\end{center}
\end{figure}

In Fig.~\ref{simulacion} (a) and (b) we plot $W(t)$ for the SRM as function of
$t$ for $\lambda=2.5$ and $\lambda=3.5$ respectively. In both
figures we can see a very short growing regime for $W(t)$ after
which the system saturates with a width $W_s$. This fast regime before
the saturation can be explained in terms of finite size effects. For
almost all growth processes the correlation length grows with time until
it reaches the characteristic length of the system \cite{barabasi},
which for complex random networks is the diameter $d$. As explained
above, the diameter is very small, and the system reaches the
saturation time very fast. We focus the attention on the steady
state of $W_s$ because only at the steady state matters to analyze
the fluctuations in the load balance of multiprocessors in parallel
computing or synchronization of queues.

For $\lambda=2.5$ we found by a
linear fitting of $W(t)$ in the steady state that $W_s$ behaves with 
$N$ as $W_s \sim \ln N$ (see the inset of Fig.~\ref{simulacion}(a)).
The same scaling behavior was obtained for all other values of
$\lambda < 3$. We also run all the simulations for an initial flat
interface and found no differences in $W_s$ \cite{nota1}.
For $\lambda=3.5$, $W_s$ does depend weakly on the system size for big
enough networks (see the inset of the Fig.~\ref{simulacion}(b)). Korniss
reported this lack of finite size effect for the growing network model of Barab\'asi-Albert \cite{rmpBarabasi} that has
$\lambda=3$ \cite{Korniss07}. 

\begin{figure}
\begin{center}
\includegraphics[width=8cm,height=10cm,angle=-90]{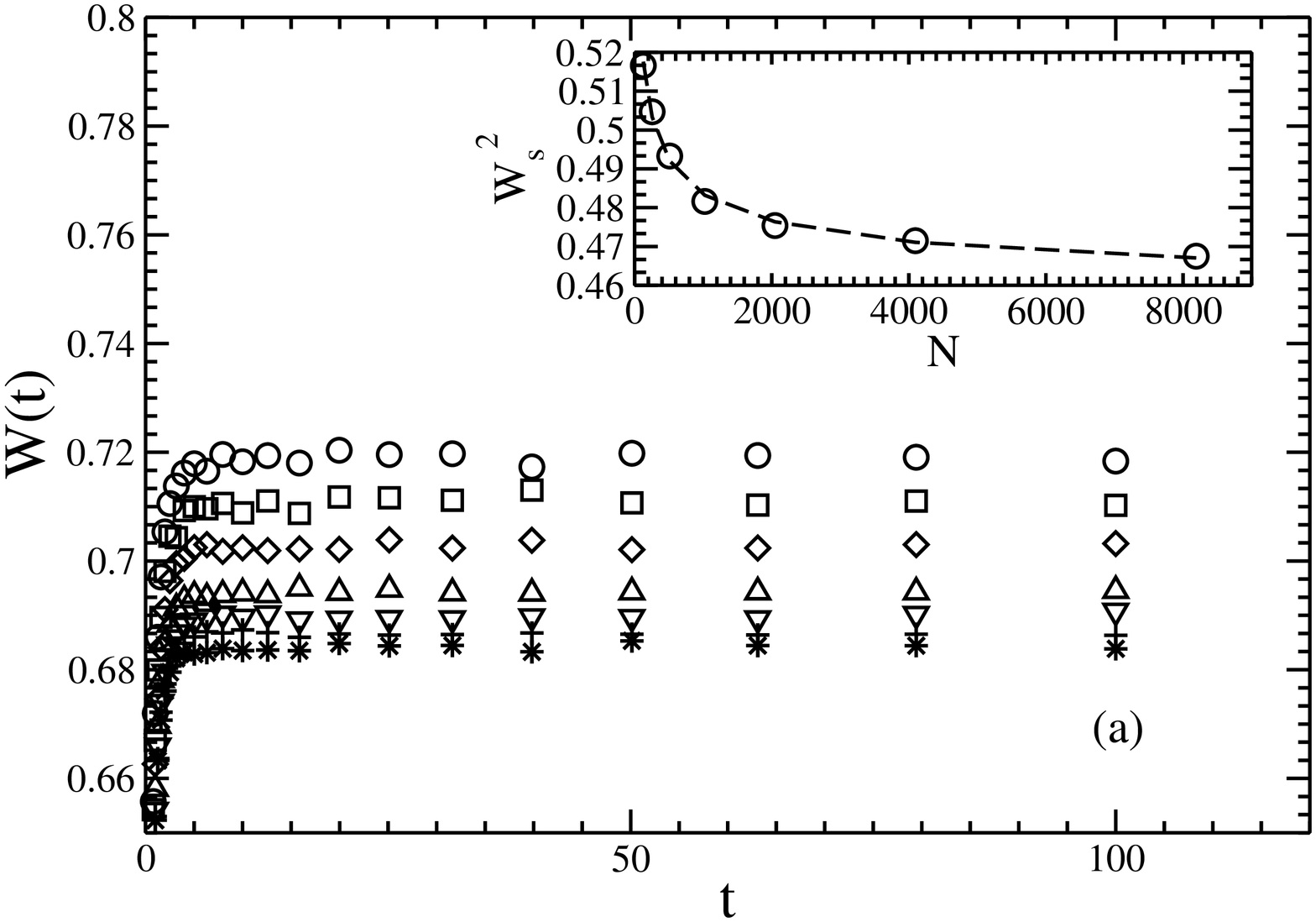}
\includegraphics[width=8cm,height=10cm,angle=-90]{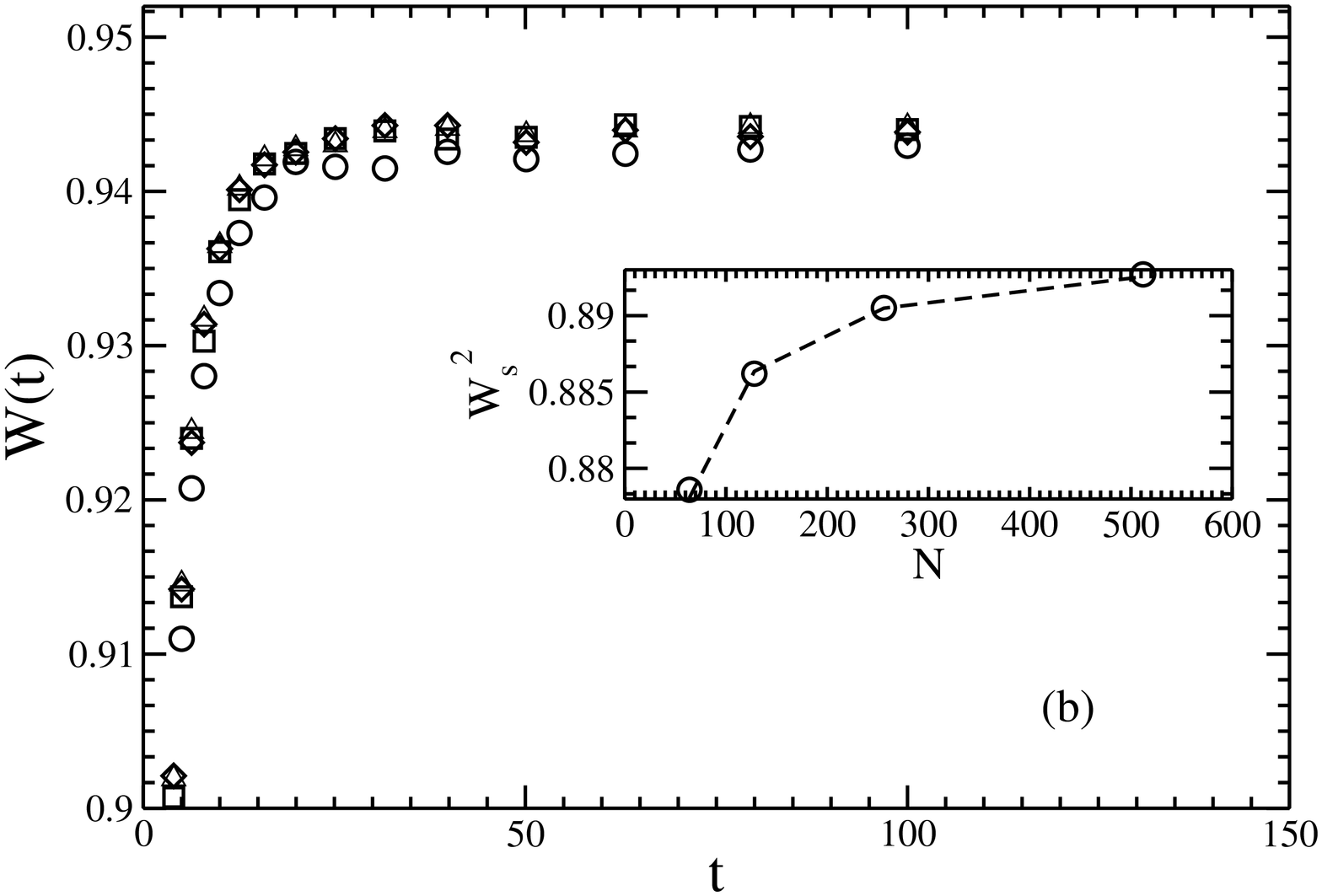}
\caption{Plots of $W(t)$ from the integration of the EW equation and
  differents system size $N$.(a) For $\lambda=2.5$, $N=128$ ($\bigcirc$),
  $N=256$ ($\Box$), $N=512$ ($\diamond$), $N=1024$ ($\bigtriangleup$),
  $N=2048$ ($\bigtriangledown$), $N=4096$ (x) and $N=8192$ (*).  We can
  see that $W_s$ decreases with $N$. In the inset figure we show in symbols
  $W_s^2$ as function of $N$.  The dashed line is the fitting with the
  Eq.(\ref{Eq.wsn}) ($A\approx 0.10 $ and $B\approx 0.75 $).  (b) For
  $\lambda=3.5$, $N=128$ ($\bigcirc$), $N=256$ ($\Box$), $N=512$
  ($\diamond$), $N=1024$ ($\bigtriangleup$). The dashed line represent the
  fitting with Eq.(\ref{Eq.wsn}). ($A\approx 1.15$ and $B\approx
  0$)\label{integracion}}
\end{center}
\end{figure}

As mentioned above, it is well known that this model in Euclidean
lattices belongs to the EW universality class represented by
Eq. (\ref{Eq.ew}), so it is expected that Eq.~(\ref{Eq.ew}) will
show the same scaling behavior as the SRM model. In this work we
demonstrate that surprisingly this is not generally true. In
Fig.~\ref{integracion} (a) and (b) we show $W(t)$ as function of $t$  
for different values of $N$ from the numerical integration of
Eq.~(\ref{Eq.ew}) with $\nu=1$ and $D=1 $ for SF networks with $k_{min}=2$
for $\lambda=2.5$ and $\lambda=3.5$.

Counterintuitive, for $\lambda=2.5$  $W_s$ decreases with the system
size, which  is a non expected result for any growth model. If this
were the case, increasing the system size will be a simple strategy to minimize
the roughness and thus improving synchronization of queues or balance in the
load of multiprocessors in parallel computing.

We next show that the decreasing of the width for $\lambda<3$ is mainly due to finite size
effects introduced by the MR construction. It was shown  in \cite{Korniss07} that for the EW process
in unweighted networks the absolute lower bound of the $W_s^2$ is 

\begin{equation}\label{eq.lowerbound} 
W^2_{min} =(1-1/N)^2 \frac{1}{\langle k \rangle}\;.
\end{equation}

The decreasing on the width observed in our numerical results is because
$\langle k \rangle$ increases with $N$. As a consequence of the 
 MR construction which introduces  the natural cutoff
$k_{max}=k_{min} N^{1/(\lambda-1)}$~\cite{pastor2004}, 
$\langle k(N) \rangle$ is given by

\begin{equation}\label{eq.kN} 
\langle k(N)\rangle = k_\infty \frac{1 -
1/N^{(\lambda-2)/(\lambda-1)}}{1-1/N},
\end{equation}

where $ k_\infty \equiv k(N \to \infty)$.
Taking into account the results presented in Eq~(\ref{eq.lowerbound}) where we replace $<k>$ by $<k(N)>$ we propose that
\begin{equation}\label{Eq.wsn}
W^2_s \sim W^2_{s}(\infty)
\left(1-\frac{A}{N}+\frac{B}{N^{(\lambda-2)/(\lambda-1)}}\right),
\end{equation}
where $W_s(\infty) \equiv W_s(N \to \infty)$.

The fitting of $W_s^2$ with Eq.(\ref{Eq.wsn}) shows an excellent agreement 
[see the inset of Fig.\ref{integracion}(a)] with the simulations supporting
that the decrease in the width for $\lambda<3$ is mainly due to the MR construction and for large $N$, $W^2_s\sim cte$.

Thus, the scaling behavior of $W_s$
for the SRM model with $\lambda <3$ is not well
represented by the EW equation with constant
coefficients $\nu$ and $D$, despite the fact that it is often used in
synchronization problems.

Next we analyze finite size effects for $\lambda>3$. For the SRM model $W_s$
was well fitted by Eq. (\ref{Eq.wsn}) [see the inset in
Fig. \ref{simulacion}(b)]. Thus, in this regime, the finite size effects can
be attributed to the MR construction. For the EW equation we find the best
fitting $W_s^2\sim W^2_s(\infty)(1-A/N)$ with $B \approx 0$ in
Eq~.\ref{Eq.wsn}.  This behavior cannot be explained as finite size effects
due the MR construction because for $\lambda >3$, $N$ diverges faster than
$N^{(\lambda-2)/(\lambda-1)}$ and may be is due to the EW process. The
separation between the finite size effect of the EW process and the MR
construction is still an open question that goes beyond the aim of this paper and could be the subject of
future researches.

In summary, we simulate the SRM model in SF networks and compare the results
with the EW process. We show that a discrete model and a continuous model
which share the same scaling properties on Euclidean lattices does not
exhibit this equivalence on complex networks.  For the SRM model in SF
networks $W_s$ diverges with the system size as $\ln N$ for $\lambda<3$.  For
$\lambda>3$ for both, the model and the EW equation, when $N\to \infty$, $W_s
\to cte$.  In order to compare the results of the SRM model with a continuous
equation further investigation including higher order of the Laplacian in the
continuous equation are needed. Also the dynamics could introduce some
weigths on the links on the underlying unweighted network that even at a
linear approximation could affect the EW unweighted process. This is the aim
of our future research.
Finally, we can conclude that despite the fact that the SRM model and the EW equation belongs to 
the same universality class in Euclidean networks, in SF networks they do not have the same behaviour.

\begin{acknowledgments}
We thanks to C. E. La Rocca for useful discussions.
This work has been supported by UNMdP and FONCyT (PICT 2005/32353).
PAM is also member of the Consejo Nacional de Investigaciones
Cient\'{\i}ficas y T\'{e}cnicas (CONICET), Argentina.
\end{acknowledgments}

\break

\break
\newpage

\begin{thebibliography}{99}
\bibitem{nota1} {If the initial
configuration is flat the second rule of Eq.~(\ref{Eq.rules}) is: if
there are $m \in v$ minimum and $j$ is one of them $ h_j = h_j+1$
with probability $1 /m$}.
\end{thebibliography}
\end{document}